# Retina gap junctions support the robust perception by warping neural representational geometries along the visual hierarchy

Yang Yue, Shenjian Zhang, Yonghong Tian, Kai Du, Tiejun Huang
School of Computer Science, Peking University, China

# Abstract

Deep Neural Networks (DNNs) are vulnerable to elaborately designed adversarial noise, although they have achieved extraordinary success in many tasks. Compared with DNNs, the human visual system is highly robust. However, it is unclear how the human visual system defends against adversarial attacks, especially the role of the early visual system and its influence on the brain manifold. Due to retina gap junctions being crucial for the denoising function in the early visual system, we combine a retina gap junction-based filter, G-filter, with DNN as an abstract human visual system model called the biological hybrid model. We adopt this model to study the defense performance of retina gap junctions and their impact on the brain manifold. Compared with other defense methods, the biological hybrid model is more robust and can be further improved by introducing noise during training. Next, we analyze the manifold and its decision boundary of the biological hybrid model from a geometry perspective. The results show that the biological hybrid model has a unique 2D decision boundary with high nonlinearity and a lower curvature of the decision boundary of the manifold compared to other defense methods. The transforming manifold may account for the high robustness of the biological hybrid model. Finally, to dissect G-filter and clarify its internal mechanism, we borrow the Neural Ordinary Differential Equation (ODE) concept and rewrite G-filter into an equivalent recurrent neural network. The results show that the decision boundary of the model's manifold will gradually change with time and eventually reach a steady state, which is modulated by gap junction conductance, revealing the influence of retina gap junctions on the brain manifold is a gradually evolving process.

# 1. Introduction

A fundamental question in systems neuroscience is how different sensory processing stages are orchestrated to produce robust perceptual behaviors. As the first stage of visual processing, the retina is an essential part of the biological visual system. The retina has been shown to contain rich circuit structures and perform complex visual computations. For example, as electrical synapses, retina gap junctions allow bi-directional and unattenuated currents to pass through between adjacent cells in order to smooth visual signals and perform blind denoising [12-14]. Retina gap junctions also enhance the signal-to-noise ratio of visual processing by flexibly adapting to diverse ambient light conditions. These findings lead to the predominant view that the retina is mainly responsible for low-level visual processing (e.g., image denoising). Its contributions to high-level visual functions remain largely underexplored.

In a parallel line of research, modern artificial visual systems, such as deep neural networks (DNNs), have achieved extraordinary performance in a variety of machine vision tasks, such as object recognition, object detection, image segmentation, and video understanding , etc. Despite the dazzling success, DNNs still fall short in several aspects as compared to human vision. One potent example is adversarial images, in which small variation in pixel intensity poses minimal impact on human vision but strongly biases object recognition in DNNs. The striking properties of adversarial images highlight the fundamental distinctions that still exist between DNNs and human vision. Adversarial images also impose severe safety concerns on machine learning applications, because deliberate image manipulations (i.e., adversarial attacks) can induce systematic failures of DNNs. The defence against adversarial attacks therefore becomes an active research field in machine learning .

The superior resilience of human vision against adversarial attacks has attracted wide research interests in developing brain-inspired algorithms for adversarial defence. Modern DNNs incorporate several mathematical operations (e.g., convolution, pooling) that bear strong resemblances to neural processing in the primate visual cortex. However, the functionalities (e.g., gap junctions) performed by the retina are largely ignored in current DNNs. This raises an intriguing question: if cortex-like mechanisms are well sufficient for accomplishing complex visual tasks (e.g., object recognition), why should the rich structures and functions of the retina be developed in the first place as the consequences of biological evolution? The example of adversarial attacks demonstrates that subtle defects in low-level image properties may be erroneously magnified by subsequent non-linear hierarchical processing and produce drastic effects on high-level visual representations. Thus, the preprocessing (e.g., denoising) performed by the retina is critical as its output acts as the basis for all downstream processing stages along the visual hierarchy. However, it is almost intractable to experimentally investigate the retina's contributions to high-level visual functions *in vivo*, because it requires (1) simultaneous recording of all neural activity along the visual hierarchy, and (2) testing a large number of stimulus conditions. These technical barriers hinder us from gaining a complete picture of the retina's function in visual processing.

In this work, we show that the retina acts as a powerful preprocessing filter to attenuate adversarial noises and support robust high-level visual representations, thereby substantially facilitating defence against adversarial attacks. To circumvent costly neurophysiological experiments, we leverage modern DNN architectures to emulate the primate visual system. DNNs have been used to investigate the neural mechanisms of a wide range of visual phenomena in primates, such as object recognition , face perception , visual adaptation , and perceptual learning , etc. Unlike the biological neural activity that is difficult to empirically assess, all neural activity and synaptic weights in DNNs are readily accessible and measurable. DNNs therefore provide an efficient and economic test-bed to shed light on neuroscientific theories.

In particular, we design a simple computational block based on retina gap junctions, dubbed G-filter, and incorporate the G-filter as a preprocessing step into several well-established DNN baseline networks. Importantly, the design and parameters of the G-filter are derived directly from our biological knowledge of the retina with no additional performance-driven optimization. Surprisingly, the brain-inspired, parameter-free G-filter outperforms other preprocessing defence methods and effectively improve defence performance. To understand the effects of the G-filter on high-level visual representations, we further assess the high-dimensional representational manifolds of object categories in DNNs. We found that the G-filter shapes the downstream neural representations to constitute a closed, trench-like ground-truth region. Interestingly, the G-filter improves the robustness of by deepening the "trench" such that adversarial perturbations can rarely pull the embedding of an adversarial image out of the trench. This unique property of the G-filter stands in stark contrast to other popular defence methods that merely focus on broadening the ground-truth regions. Furthermore, we rewrite the G-filter into an 600-layer recurrent neural network called "ShallowRetinaBlock (SRBlock).", and characterize the spatial (i.e., layer-by-layer) and temporal (i.e., step-by-step) properties of the network. The results show that the unique decision boundaries induced by the G-filter gradually evolve and eventually reach a steady-state shape. We also find that these advantages of gap junctions disappear if the gradient during network training is allowed to penetrate SRBlock, because the backpropagation mechanism leaves the leeway for the adversarial attacks targeting training gradient. This finding also suggests that the lack of feedback projections from cortex to retina in the biological visual system may reflect the optimal balance with respect to the stability-plasticity dilemma—that is, higher-level visual stages remain plastic in order to confront the ever-changing environment but lower-level front-end filters keep stable in order to maintain the statistical homeostasis of early visual signals.

In sum, our work leverages modern DNNs to study the functional role of the retina under the context of the whole visual hierarchy. Our results demonstrate that a key function of the retina is to strengthen the robustness of high-level visual representations and facilitate adversarial defense. We design a parameter-free computational block purely based on biological knowledge of the retina. This off-the-shelf solution from neuroscience shows competitive performance in adversarial defense and exhibits novel defense mechanisms that differ from existing methods. We also provide a differentiable, ODE-based version that is plug-and-play for many modern DDN architectures. The results here reveal the deep connections between the biological and artificial visual systems, and embody the long-hypothesized view that brain-inspired mechanisms can be utilized to promote defense against adversarial attacks, and more broadly, machine learning applications.

# 2. Results

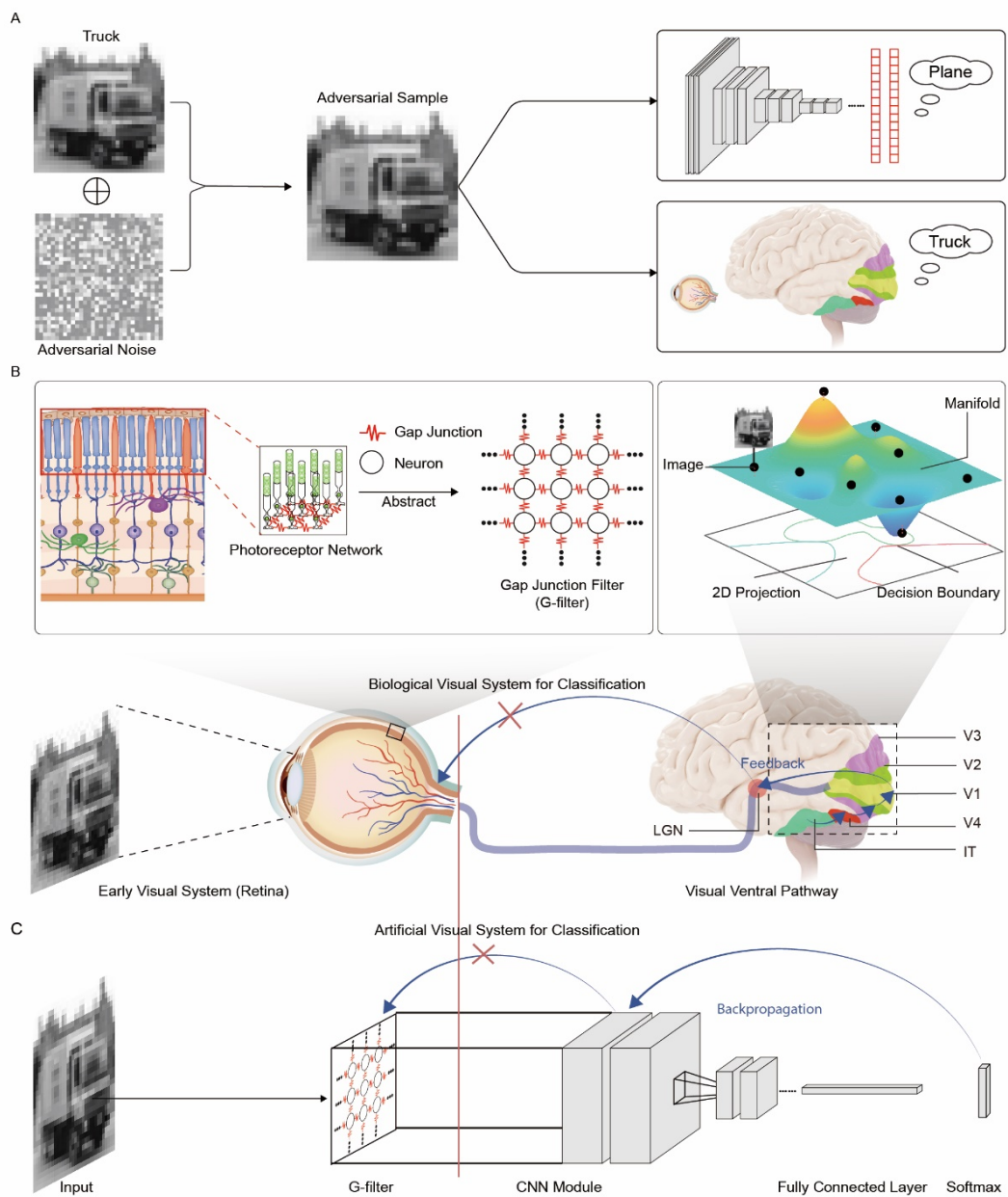

**Figure 1: Retina gap junctions support the robust perception by warping the manifold along the visual hierarchy.**

(A) The artificial neural network is significantly less robust against adversarial noise than the human visual system due to the neglect of the retina in the deep neural network architecture.

(B-C) This work adopts a G-filter derived from the retinal photoreceptor network as a retinal model and combines it with a DNN to simulate the visual hierarchy from the retina to the visual cortex. Since the retina has no feedback signal from the cortex, our model uses the G-filter as a preprocessing module whose parameters are fixed during training and testing (i.e., no error signal is back-propagated from the CNN to the G-filter). We train this hierarchical model for a classification task to study how retinal gap junctions support robust perception along the visual hierarchy through its manifold and decision boundary. The manifold is defined as the high-dimensional space coordinated by the neural activity of the last layer of the CNN. The decision boundary is the hypersurface of the neural manifold that divides the manifold into different classes. We investigate the influence of the retina on high-level cortical visual processing by analyzing how the manifold and its decision boundary change with and without G-filters.

To understand how retinal gap junctions support biologically robust perception along the visual hierarchy, we propose a biological vision model consisting of a gap junction filter (G-filter) and a deep neural network, where G-filter serves as a simplified retinal model, DNN as a ventral visual pathway model. First, we experimentally demonstrate that G-filter can significantly improve the robustness of various popular DNN architectures to various adversarial attacks in image classification tasks. Next, we analyze the manifold of the biological vision model and its decision boundary by means of neural population geometry to understand how the model produces robust perception. The results demonstrate that G-filters support robust perception by warping the geometry of neural representations along the visual hierarchy. Finally, to improve computational efficiency and analyze how the warped neural representation geometry evolves, we convert the G-filter to a recurrent neural network and combine it with a DNN as an end-to-end neural network.

# Retina gap junctions support the robust perception along the visual hierarchy

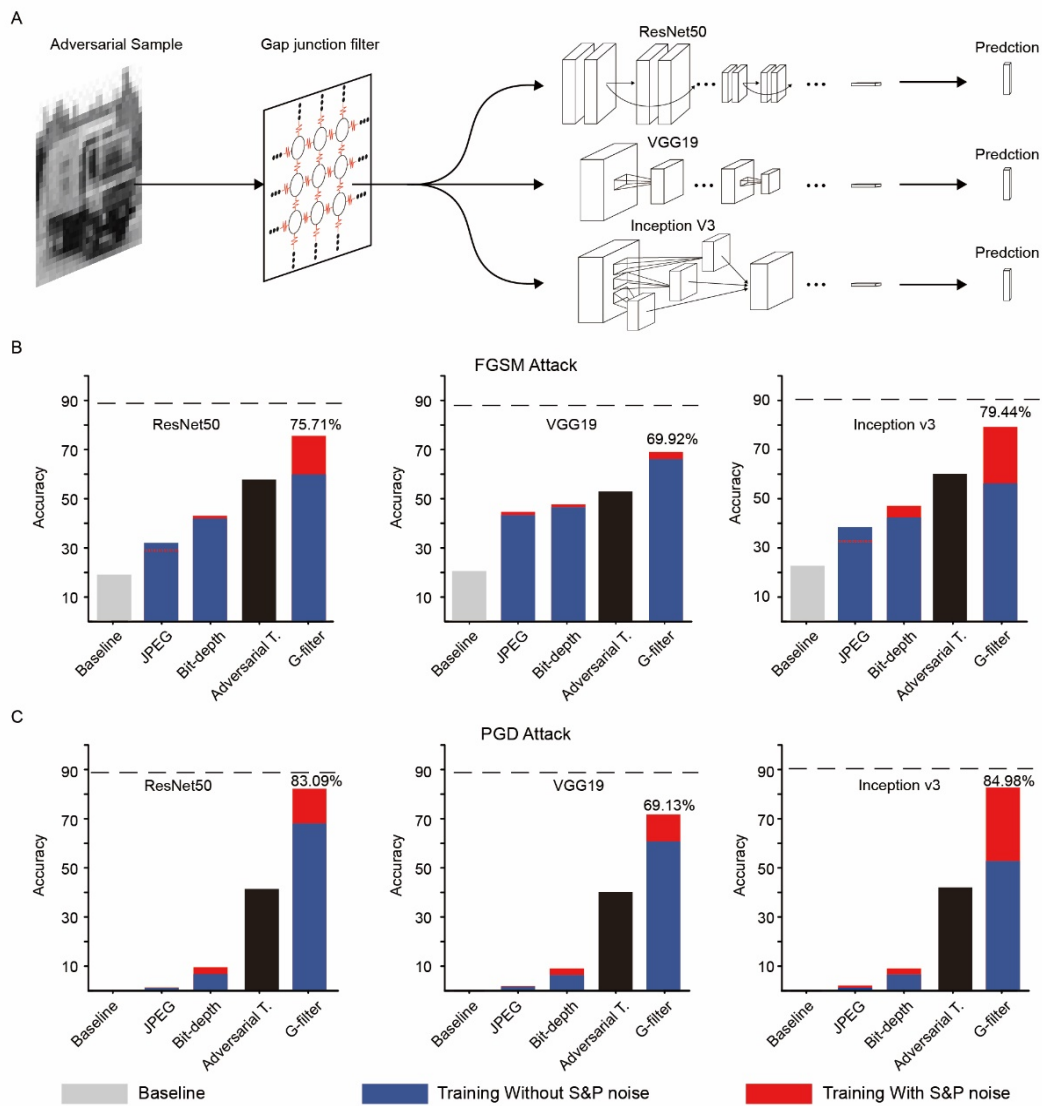

**Figure 2: The gap junction filter improves the DNNs' robustness to adversarial attacks.**

**(A)** The architecture of the biological vision model. Each input image is preprocessed with the G-filter and then feed into three representative CNN architectures: ResNet50, VGG19, and Inception V3. The parameters of the G-filter are fixed during training and testing.

**(B-C)** The performance of the three architectures under FGSM and PGD attacks. The biological vision model (the G-filter preprocessed CNNs) outperforms other preprocessing defenses such as JPEG compression, bit depth reduction, and adversarial training under FGSM and PGD attacks (black bars for adversarial training and red bars for others). The performance of the biological vision model under adversarial attack can be further improved by training with Salt & Pepper (S&P) noise, which is illustrated by red bars in the diagram.

To demonstrate that retina gap junctions support robust perception along the visual hierarchy, we construct a biological vision model that incorporates the retina and ventral visual pathways (Fig. 1C).

The retina model comes from our previous work[54] and we model the retina photoreceptor network as a conductance-based grid network, called G-filter. The G-filter is considered as a simplified model of the retina, where the photoreceptors are connected by gap junctions and the network size is equal to the size of the input image.

Specifically, neuron $i$ of the G-filter receives a photon stimulus $I_i^{photon}$ converted from the pixel value $x_i$ of the input image $x$ (Methods). Besides, each neuron receives currents $I_{ij}^{gap}$ from neighboring neurons via gap junctions, which is determined by the difference between the membrane potentials of the two neurons and the conductance $g_{gap}$ of the gap junction, given by equation (2) (Methods). In this study, we consider all gap junctions to have equal conductance. Taken together, the dynamics of the membrane potential of neuron $i$ is given by equation (1).

$$c_m \frac{dV_i}{dt} = -g_L(V_i - E_L) + \sum_j I_{ij}^{gap} + I_i^{photo} \quad (1)$$

$$I_{ij}^{gap} = g_{gap}(V_j - V_i) \quad (2)$$

where $c_m$ is the membrane capacitance, $g_L$ and $E_L$ are the conductance and reversal potential of the leak channel respectively. The dynamics of the G-filter is simulated for 300 ms for each input image.

Next, we adopt a max-pooling layer to select the maximum value of the absolute value of the membrane potential of each neuron within the 300 ms simulation window, and then normalize these voltage values as input to the CNNs (Figure 2A).
Considering the origin of convolutional neural networks from the ventral visual pathway , we select three representative CNN architectures that have achieved great success in various intelligence tasks: ResNet50[1], VGG19[30], Inception-V3[31] as the ventral visual pathway models (Fig. 2A).

Finally, we train our biological vision model (see Methods) via backpropagation . It is worth noting that only the parameters of CNNs are tunable, while the parameters of G-filter (i.e. $g_{gap}$) are fixed throughout the training procedure. This setting follows the biology, where there is no feedback pathway (i.e., feedback error signals) from the visual cortex to the retina.

To test whether G-filter can improve the robustness of the three CNNs, we employ two essentially different adversarial attack methods: Fast Gradient Sign Method (FGSM) and Projected Gradient Descent (PGD) attack . FGSM is an efficient but mild one-step attack method, while PGD is a more powerful iterative attack method. In general, PGD attacks are more effective than FGSM, because the baseline performance (the performance of CNNs without any defense method) under PGD attack drops more than FGSM.

Since G-filter in the biological vision model is a preprocessing module, we first compare the performance of G-filter under adversarial attack with the classical preprocessing defense methods: JPEG compression and bit width reduction . The results show that under different architectures and attack methods, G-filter consistently and significantly outperforms preprocessing defense methods (performance drops to ~20%) (Fig. 2B&C). The performance of three biological vision models (G-filter combined with ResNet50, VGG19 and Inception V3) is 56.94%, 70.89%, and 56.04%, respectively. Interestingly, the advantage of G-filter is more prominent under PGD attack (Fig. 2C), where the preprocessing defense method drops to about 10%, while the performance of the three biological vision models is 47.52%, 68.41%, and 43.61%, respectively.

Considering that the retina receive stochastic noise from the environment, can the robustness be further improved if stochasticity is introduced during training, i.e. training a biological vision model with noisy images. The results show that training with noisy images can improve the robustness of biological vision models, where training with salt and pepper (S&P) noise has the greatest effect. Surprisingly, the two preprocessing methods show little or even decrease in robustness under adversarial attacks after training with S&P noise, while the biological vision models trained with S&P noise significantly and reliably improve the performance with an average performance improvement of about 20% (Figure 2B&C).

We then compare the performance of the biological vision models with another noise–based defense method, the adversarial training method, which iteratively adds adversarial noise to training images during training. The results show that adversarial training outperforms the JPEG and Bit-Width methods because it "saw" adversarial examples during training. Interestingly, the defense performance of the biological vision models trained with S&P noise is close to the performance of the baseline (~89%), where the performance of the three biological vision models is 75.71%, 69.92%, and 79.44% under the FGSM attack (Fig. 2B; red); 83.09%, 69.13%, 84.98% under PGD attack (Fig. 2C; red).

In summary, our experimental results show that retinal gap junctions play an important role in maintaining robust visual perception, and the introduction of stochasticity can further improve robust perception. Interestingly, the introduction of stochasticity is only effective in the bionic model and cannot improve the robustness of the two preconditioned defense methods.

# Retina Gap Junctions warp DNN's 2D neural representation geometries

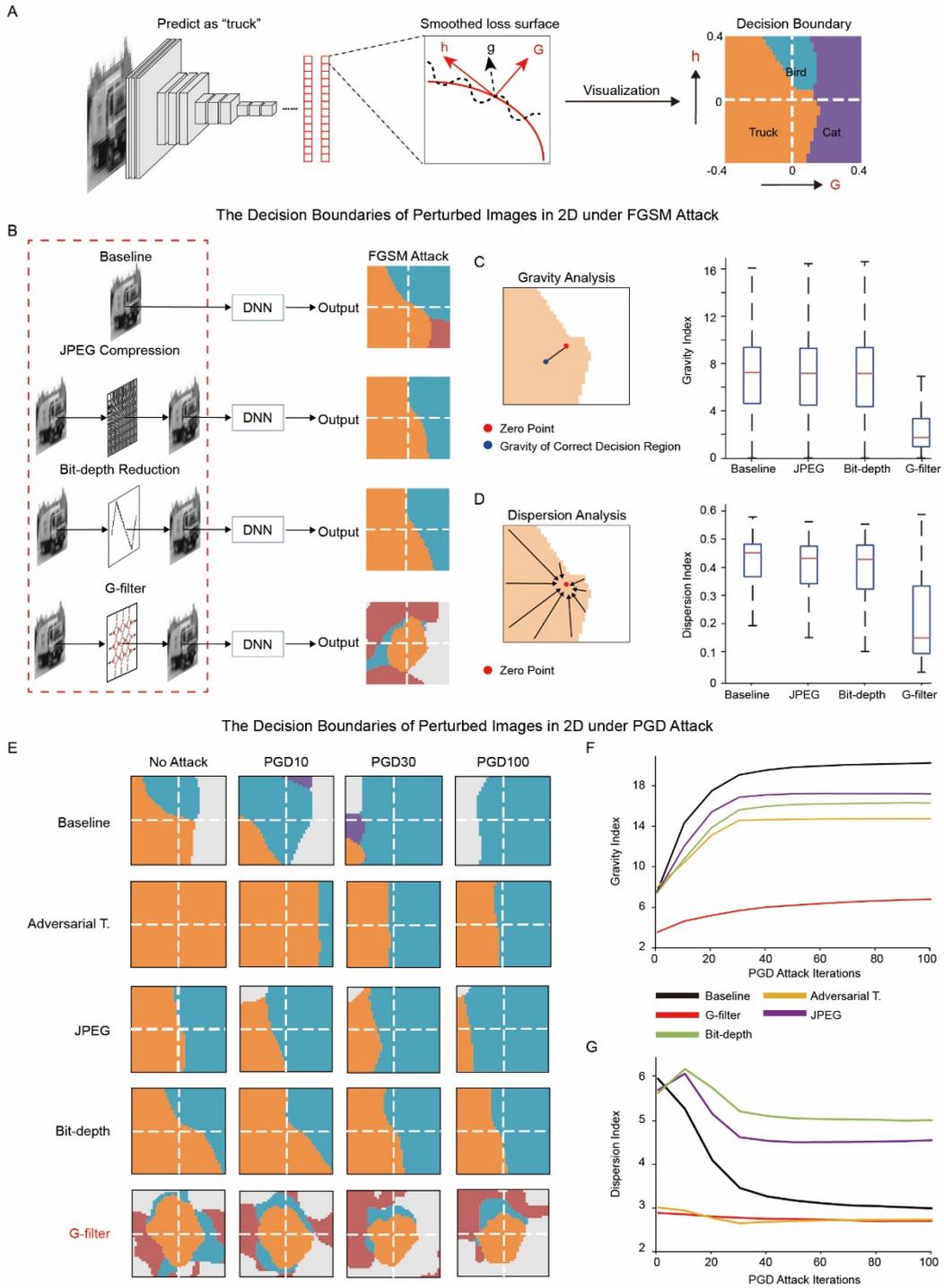

**Figure 3: The G-filter transforms DNN's 2D decision boundary into a unique circle-like shape.**

**(A)** The visualization of CNN's manifold and its decision boundary (Resnet50). We use the smooth gradient G of the network and a vector h perpendicular to it to project the original gradient g on a 2D plane. The center point of the plane is the input image, and the other points are the positions of the perturbed image. Different colors represent different categories, and the dividing line between them is the decision boundary.

**(B)** and **(E)** 2D decision boundary formed by different defense methods before attack and after PGD attack, respectively.

**(C-D)** and **(F-G)** Quantitative analysis of the distance from the center of gravity of the decision boundary to the center point and the dispersion of the decision boundary to the center point for different defense methods before attack and after PGD attack, respectively.

Although we experimentally verified that retina gap junctions play an important role in robust perception, the mechanism underlying the robustness endowed by gap junctions remains unknown. Next, we employ the neural population geometry to understand the mechanisms of biological vision models for robust perception. This method analyzes the geometry of the high-dimensional representation of the input information by leveraging mathematical and computational tools [76].

Here, we extended previous approaches to investigate the robustness mechanisms by visualizing and quantifying the geometry of decision boundaries of the CNN (using Resnet50, the same below) after training. We projected the input subspace formed by each image onto a 2D square, whose origin $(0,0)$ is the input image (Figure 3A). The direction of the horizontal axis is the smooth gradient $G$, while the direction of the vertical axis is a random direction $h$ orthogonal to $G$. (Figure 3A; Methods). Therefore, each point $(a,b)$ in the visualized 2D square represents an image perturbed from the original image by perturbation $aG + bh$ (Figure 3A; Methods). The entire 2D square projects the manifold and its decision boundary near the input image. The smoothed gradient is computed by averaging 10 random samples of the original gradient. The smooth gradient G removes the local non-informative noise of the gradient to make it more representative. The classification result of each perturbed image in the 2D square is color-coded in the visualization (Figure 3A). The curves that distinguish different classes are decision boundaries.

First, we plotted 2D squares of truck images for the baseline CNN and the CNN trained using JPEG, Bit-Width, adversarial training, and G-filter (Figure 3B). Interestingly, we found that only CNNs trained with G-filters formed a unique circle-like decision boundary (Fig. 3B). We then quantified two geometric properties of the decision boundary for all test images: gravity analysis and dispersion analysis (Fig. 3C&D; Methods). The results show that DNNs trained with G-filter are significantly different from other methods in two metrics (Figure 3C&D).

Next, we investigated how the decision boundary changed under the iterative PGD attack (Figure 3E). For the baseline DNN and two preprocessing methods, the original image (i.e., without attack) was very close to the decision boundary (Figure 3E). For the adversarial training method, although the correct decision region without attack covers the entire 2D square, the correct decision region shifts rapidly with PGD attack (Fig. 3E). However, for the G-filter, the shape of the decision boundary remained relatively stable over 100 iterations (Figure 3E). We then quantified the gravity index and dispersion index for all test images. The two quantities of the G-filter remained stable and small within 100 iterations, while the quantities of the other methods increased significantly after a small number of iterations (Fig. 3F&G). The change of gravity index shows the shift of the decision boundary during the iteration, and the change of the dispersion index shows the uniformity of the decision boundary during the iteration. Combining these two indexes, the unique decision boundary formed by G-filter has small offset and uniform shape under iterative attack. Thus, under 2D projection, retina gap junctions warp the 2D neural representation geometry of the CNN.

## Retina Gap Junctions warp DNN's high dimensional neural representation geometries

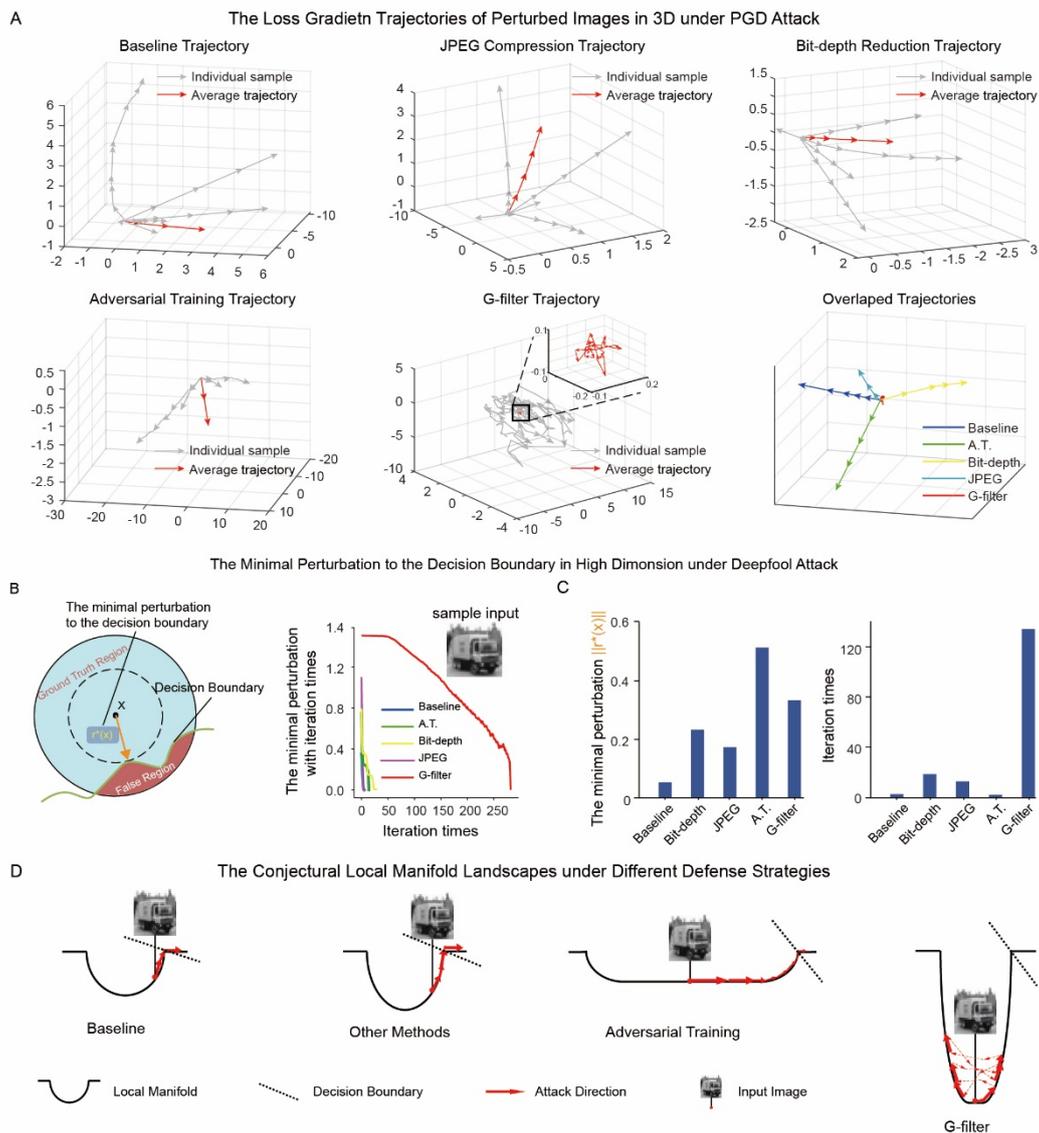

**Figure 4: The quantitative analysis of the manifold in the high-dimensional space.**

(**A**) The first five plots are visualizations of PGD attack trajectories of the CNN (Resnet50) trained with different defense methods, similar to Figure 2E but projected to a 3D space. The origin point of the plot is the original input image. Each axis is a direction chosen by LLE[48] that can preserve distances within local neighborhoods. The gray curves are trajectories of five sample images and the red curve is the average trajectories of all test images. The sixth plot is the overlap of average trajectories of 5 different defense methods.

**(B) Left**: schematic illustration of worst-case robustness analysis. Each point within the outer circle is an image and the center of the outer circle is the original input image $x$. The red and blue parts of the outer circle represent two image classes. The CNN will classify images from the blue region (Ground-Truth Region) as the ground-truth class of the image $x$, while will classify images from the red region (False Region) as another different class. The minimal perturbation $r^*(x)$ is a vector from input image $x$ to a point on the decision boundary with the smallest length. **Right**: the current minimal perturbation as a function of iteration of five different defense methods of the truck image.

**(C) Left**: the length of the minimal perturbation of five different defense methods averaged over all test images. Right: the number of iterations to find the minimal perturbation of five different defense methods averaged over all test images.

**(F)** The conjectural 2D structures of local manifolds and attack trajectories of baseline and different defense methods.

Since the G-filter warps the 2D neural representation geometry of the CNN, showing unique geometric features, the 3D neural representation geometry may also be affected. Therefore, we next visualized the PGD attack trajectory in 3D via Local Linear Embedding (LLE) (Fig. 4A; Methods). The PGD attack trajectories of other defense methods are similar to straight lines (Figure 4A), indicating that the PGD method can accurately identify the most effective attack directions. However, the PGD attack trajectory of the G-filter is restricted to a local region (Fig. 4A), making it difficult to determine the optimal attack direction. Furthermore, the average size of the PGD trajectories for the G-filter is much smaller than other defense methods (Figure 4A bottom right). Significant differences in attack trajectories between G-filter and other methods suggest that retina gap junctions warp the CNN's 3D neural representation geometry.

Further, we investigate the effect of the G-filter on the geometry of high-dimensional neural representations of the CNN. Fawzi et al. adopt the Deepfool attack to iteratively find the smallest perturbation that misclassifies an image to evaluate the geometric properties of decision boundaries for the high-dimensional manifold. The minimum perturbation is the vector $r^*(x)$ from the input image x to the decision boundary with the minimum length (Fig. 4B; Methods).

We leverage this method to calculate how many iterations are required to find the minimum perturbation of the baseline and the four defense methods. For the example truck image in Figure 4B, the Deepfool attack needs over 12 times more iteration times to find the minimal perturbation of G-filter compared to other defense methods (Figure 4E). On average across all test images, although G-filter has the second to the largest minimal perturbation (Figure 4D), it takes more than 10 times more iteration times to find the minimal perturbation (Figure 4E). These quantifications suggest that retina gap junctions warp the high-dimensional neural representation geometry of the CNN. The manifold warped by the G-filter slightly reduces the length of the minimum perturbation, which in turn exponentially increases the difficulty of finding the minimum perturbation.

What is the structure of the manifold that has such an effect? From the previous analysis, it can be seen that the projection of the biological vision model in the iterative attack in 2D will form a circle-like stable decision boundary (Figure 3E-F), and the attack trajectory in the 3D projection will be limited to a small region. (Fig. 4A), a large number of iterations are required in high-dimension to arrive at the decision boundary (Fig. 4C). Taken together, we assume that the G-filter warps the manifold of the CNN and forms a deep well-like structure (Fig. 4E right) and the input image resides at the bottom center of the well. As a consequence, iterative attack methods can not quickly and accurately identify the most efficient attack direction (Figure 4E right). However, for other defense methods, the input image either lies close to the decision boundary (Figure 4E left two), or the manifold is shallow, so the best attack direction stays unchanged for multiple iterations (Figure 4E middle right). This conjecture might be one possible explanation based on our quantifications, but there can be other possibilities.

**Shallow Retina Block**：a module for in-depth study of how retina gap junctions warp neural representation geometries

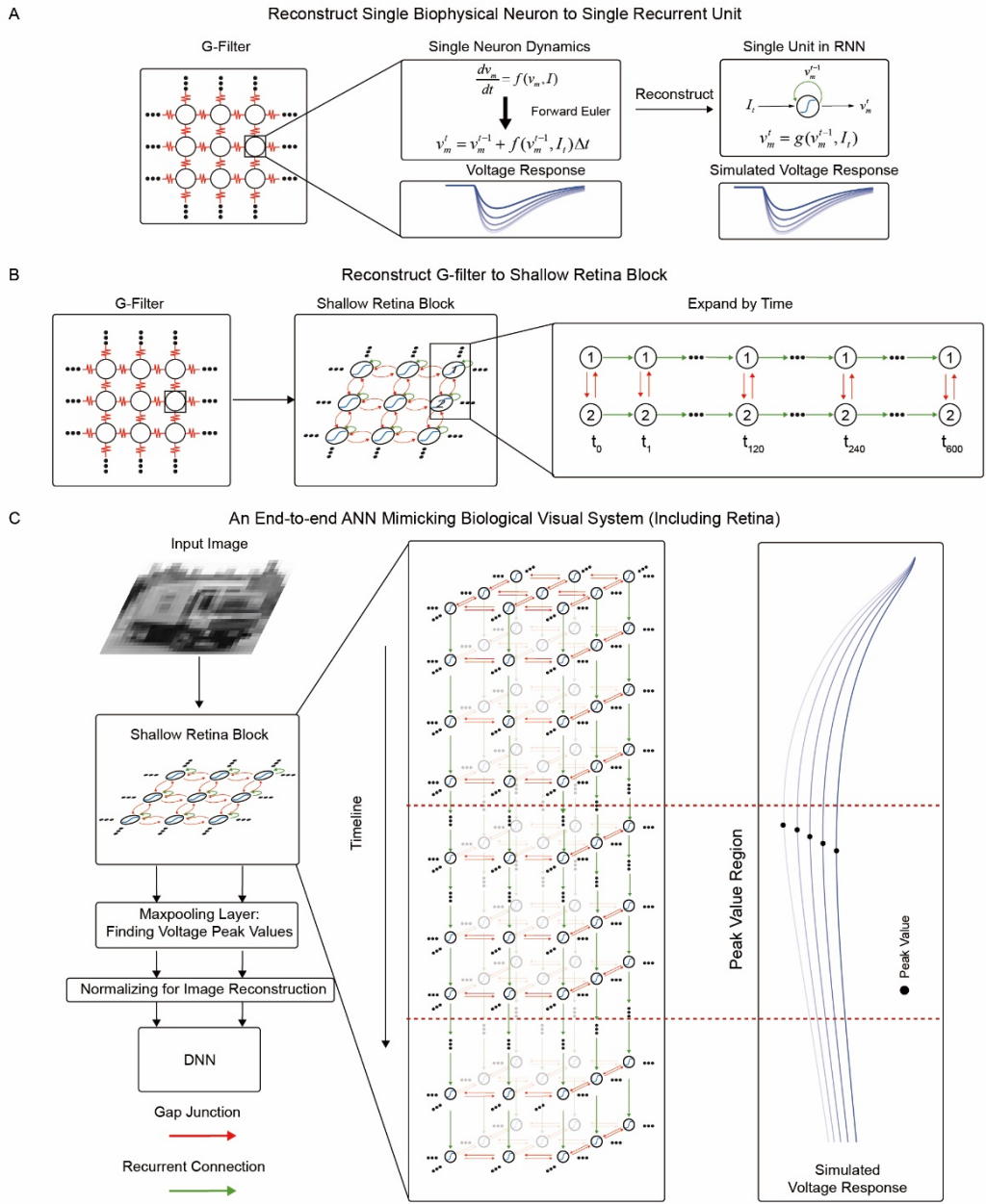

**Figure 5: The architecture and workflow of Shallow Retina Block.**
**(A)** Mathematical reconstruction of biophysical neurons in the G-filter to recurrent artificial neurons.
**(B)** Expand the recurrent connection of the shallow retina block by time into layers. Each layer represents a timestep.
**(C)** The entire end-to-end network architecture. The peak values of neurons in the shallow retina block are the input to the downstream DNN.

From the previous results, it can be found that retina gap junctions support robust perception by warping the neural representation geometries along the visual hierarchy. However, how retina gap junctions warp neural representation geometries (such as decision boundaries under 2D projections) remains elusive. On the other hand, current implementations of biological vision models are not end-to-end and computationally expensive, making it difficult to deeply analyze the interaction between the G-filter and CNNs. Inspired by neural ODEs, we rewrite the G-filter into an equivalent recurrent neural network, termed as Shallow Retina Block (SRBlock).

We reconstructed each biophysical neuron in the G-filter as a recurrent artificial neuron based on the dynamics of its membrane potential $V_i$, given by equation (1)-(2) (Figure 5A). We approximated the dynamics numerically by the forward Euler method (Figure 5A), given by equation (3):

$$V_i^t = V_i^{t-1} - \frac{\Delta t}{c_m}\left(g_L\left(V_i^{t-1} - E_L\right) - \sum_j g_{gap}\left(V_j^{t-1} - V_i^{t-1}\right) + I_i^{photon,t-1}\right) \quad (3)$$

where $v_i^t$ is the membrane potential of artificial neuron $i$ at time step $t$, $\Delta t$ is the numerical time step.

The recurrent connection can be expanded to layers by time (Figure 5B). Each layer corresponds to a time step of the dynamics, and every neuron in the layer outputs a voltage value (Figure 5B&C). The voltage value is passed to the next layer via the recurrent connection (Figure 5B&C). Now, the SRBlock can be integrated with the DNN as an end-to-end manner (Fig. 5C). The peak value of the voltage values over time will be calculated by the max-pooling layer and the normalization of the peak values will further serve as the input to the downstream DNN (Figure 5C). Therefore, we successfully rewrote G-filter as an artificial RNN, which can be further trained with the DNN in an end-to-end manner to investigate how the unique 2D decision boundary is formed.

# Retinal gap junctions gradually warp neural representation geometries through information integration

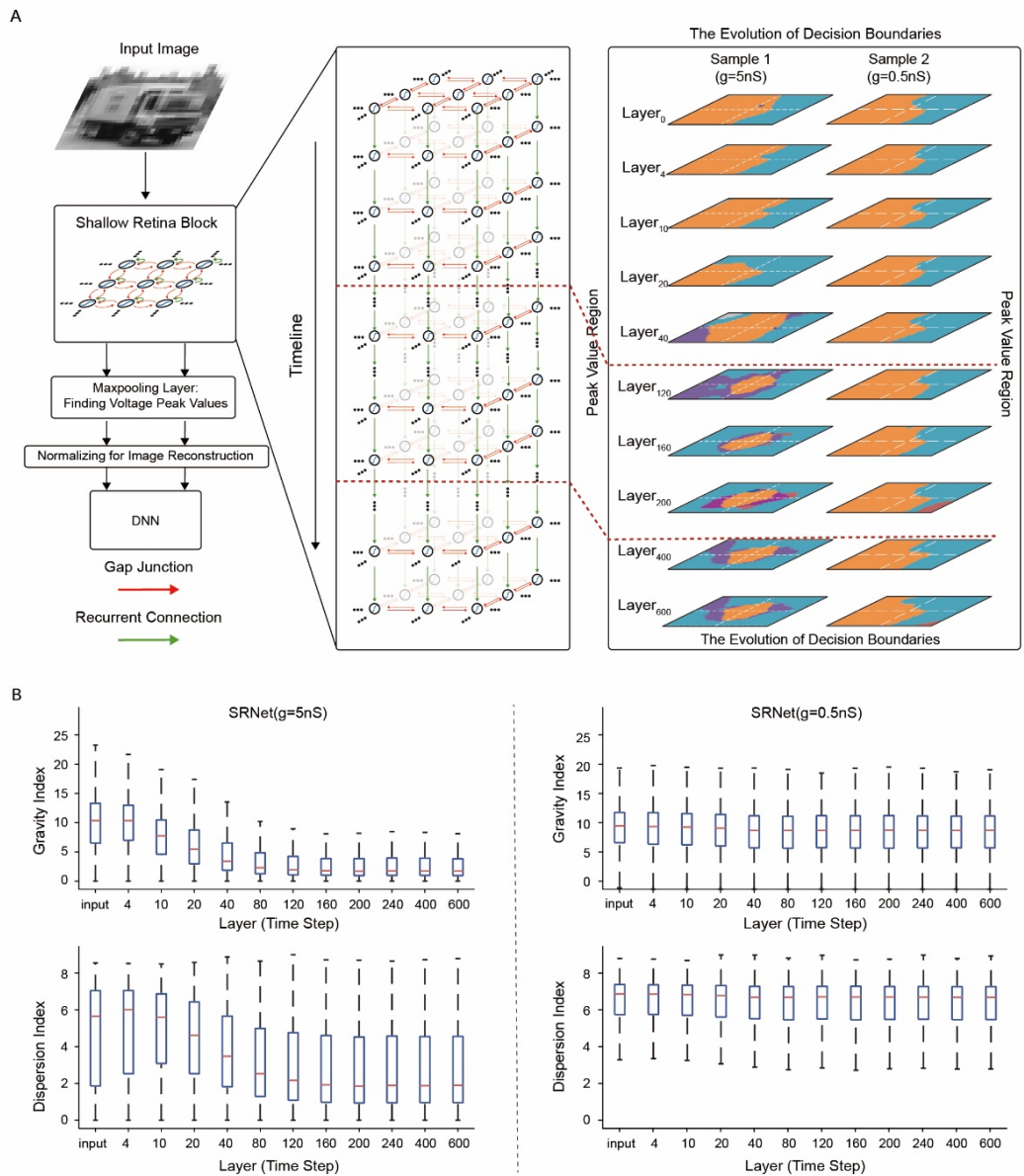

**Figure 6: The unique 2D decision boundary is gradually formed through the information integration of gap junctions.**

**(A)** We visualized the evolution of decision boundaries from two SRBlocks with the maximal gap junction conductance of 5 nS and 0.5 nS at different time steps.

**(B)** Plots of gravity index (top row) and dispersion index (bottom row) of the decision boundaries of two SRBlocks with the maximal gap junction conductance of 5 nS and 0.5 nS as a function of time steps.

After rewriting the G-filter as an SRBlock, we can explore the dynamics at each time step inside it to understand how retinal gap junctions warp the CNN's neural representation geometry. Here, SRBlock is still combined with the subsequent CNN as a preprocessing module, whose weights are fixed during training, which is indispensable to maintain the robustness of visual perception (Discussion). By expanding SRBlock by time, we obtain a neural network with 600 layers, each layer corresponding to a simulation time step of 0.5 ms and a vector of voltage values (Fig. 5C and Figure 6A).

Similar to previous analysis, we visualized the 2D decision boundary of each layer (i.e., each time step) of the recurrent neural network of the same input image (Figure 6A). For the network with gap junction conductance of 5 nS, the evolution of 2D decision boundary can be divided into two phases: formation and stabilization. In the formation stage, the 2D decision boundary gradually evolves into a unique circle-like shape from layer 0 to layer 120 (i.e., 0-60 ms; Figure 6A&B); in the stabilization stage, from layer 120 to layer 600 (i.e., 60-300 ms), the 2D decision boundary remains stable (Fig. 6A&B). It is worth noting that the voltage peak region of SRBlock also ranges from layer 120 to layer 240 (Figure 5C). The 2D decision boundary stabilizes since the peak voltage of the neuron is reached, suggesting that the peak value region is critical for encoding the low-level features of images and crucial for robustness of high-level processing. Interestingly, however, a network with a small gap junction conductance (i.e., 0.5 nS) fails to form such a manifold for most test images (Figure 6A&C), suggesting that a strong communication between retinal photoreceptors via gap junctions is essential for forming such a unique manifold and improving the robustness of high-level visual processing.

In summary, through layer-by-layer analysis, the formation of the unique circular-like 2D decision boundary evolves around 60 ms. After inputting external signals, information integration is carried out through gap junctions, and the neural representation geometry is gradually warped, so that the 2D decision boundary gradually becomes a "circle", and the centroid of the decision region is closer to the center point, thereby improving the robust perception of the biological vision model (Figure 6A & B).

## 3. Discussion

In this work, we combine retina gap junction-based G-filters with several classical machine learning architectures to slightly mimic the human visual system, called the biological hybrid model, and evaluate the impact of retina gap junctions on the robustness to adversarial attacks. G-filter significantly improves the neural network's defense performance against adversarial attacks compared with other preprocessing-based methods. Interestingly, introducing noise during training can specifically improve the performance of the biological hybrid model. Next, we evaluate the influence of retina gap junctions on the manifold from a geometrical perspective. The results show that only the 2D projection of the manifold of the biological hybrid model and its decision boundary forms a unique centralized, closed, homogeneous region. Further, we analyze the manifold and its decision boundary in the high dimension. The results show that the biological hybrid model has higher nonlinearity and lower decision boundary curvature compared to other defense methods, implying higher robustness. These results suggest that the G-filter transforms the manifold of the network, likely responsible for the higher robustness of the biological hybrid model. Finally, we borrow the concept of Neural ODE to rewrite the G-filter as a recurrent neural network called SRBlock. It allows us to dissect the G-filter to understand its inner mechanism. Noticeably, the construction process of SRBlock can be generalized to other ODE-based dynamic systems, thereby facilitating the transfer and application of biological models in machine learning. Next, we expand SRBlock into a 600-layer neural network by time step. Our layer-by-layer analysis results show that the effect of retina gap junctions on the manifold is a gradual evolution process, and the conductance of gap junctions also has an essential impact on the evolution of the manifold. Overall, the retina gap junction network defends against adversarial attacks by transforming the manifold of the neural network.

**Comparison with preprocessing-based methods in computer visions**

Current defense methods[39] can be roughly divided into two categories: preprocessing-based methods and model-based methods. Preprocessing-based methods defend against adversarial attacks by processing input images without modifying the network. Because these methods do not involve modifying the network, it is inexpensive and has great potential in practice. For example, Guo et al. propose several transformations to preprocess the input image, including bit depth reduction, JPEG compression, and total variance minimization[33]. Shaham et al. preprocess input images using low-pass filters, PCA, etc[43]. Although these methods have certain defensive effects, some are sensitive to parameters and less robust to high-strength adversarial attacks. To defend against high-intensity adversarial attacks, a complex generative adversarial network (GAN) is further introduced[44]. The input image is first sent to the GAN, and the generated new image is used as the subsequent network's input.

Compared with these defense methods, the defense performance of the G-filter against high-strength adversarial attacks is higher than other methods based on filtering or transformation. The G-filter only has one adjustable hyper-parameter - gap junction conductance, which facilitates parameter tuning. Although adopting generative adversarial networks can improve defense performance, the computational cost becomes expensive.

**Comparison with biologically inspired methods**

Research has recently drawn inspiration from neuroscience to design biologically inspired defense methods. For example, the foveal mechanism[21] derives from the function of the fovea. The fovea is a small central pit in the retina that provides high visual acuity and detail[45]. But the fovea has only two degrees of the visual field[46,47]. They adopt this mechanism to remove the image background and leave only the object as input. Then, inspired by nonlinear dendritic integration, saturating networks are propose[7]. They alter the activations in DNNs to obtain a highly kurtotic weight distribution for defending against adversarial attacks. Furthermore, to model the primate V1 cortex, Dapello et al. have developed a hybrid neural network called "VOneNets"[8].

Compared with these defense methods, the design of the G-filter is mainly inspired by the structure of retina gap junctions. The foveal mechanism or dendritic computing for defending against adversarial attacks leverages biological functions. Conversely, VOneNets improve network robustness by introducing the structure of V1. However, VOneNets do not consider the retina's possible crucial role in preprocessing. Overall, most of these biologically plausible approaches draw inspiration from the cortex, and approaches involving the retina do not focus deeply on the structure of the retina. Therefore, most defense methods probably underestimate the importance of the preprocessing mechanisms arising from the unique architecture of the retina.

**The Penetration of the SRBlock**

Since we have rewritten G-filter as SRBlock to form an end-to-end model with subsequent DNNs, gradient backpropagation can also be applied in SRBlock. Similarly, by analyzing the layer-by-layer change of the gradient during backpropagation, we find that the gradient first increases and then decreases when the gradient passed through the peak value region in backpropagation (Fig. S3A). By feeding this gradient to FGSM and PGD, we successfully attack SRBlock (Fig. S3B), indicating that gap junctions are likely to mask the true gradient through nonlinear operations to defend against adversarial attacks. For example, after penetrating SRBlock, the adversarial noise generated by both attack methods is similar to that of attacking an unprotected DNN (Fig. S3C). Further, we analyze the changes of the 2D decision boundary on two metrics after penetrating SRBlock. The results show that the unique decision boundary cannot be formed, and the effect of improving robustness is lost.

Nevertheless, there is no biological feedback from the cerebral cortex to the retina[49], and our visual system likely leverages this property to defend against environmental perturbations. In addition, SRBlock is only obtained from simplified photoreceptors and gap junctions, while the entire retina is composed of ion channels, dendrites and chemical synapses, as well as complex connections to form a much higher nonlinear system than SRBlock (Fig. S3E). According to the existence theory of adversarial examples[21,34], such a highly nonlinear system is highly robust.

## 4. Methods

### Gap junction Filter (G-filter)

Based on retina gap junction network, we have developed a G-filter for facilitating blind denoising and classification in the visual hierarchy in our previous work[23]. The G-filter only keep passive leak channel and gap junction connections, which can be described as the follows:

$$C_m \frac{dv_m}{dt} = -g_{leak}(v_m - e_{leak}) + \sum_{i=1}^{k} I_{gap}^i + I_{photo}$$

The neighboring neurons are connected by gap junction. Based on the Ohm's law, their bidirectional gap junction current $I_{gap}^i$ can be denoted as:

$$I_{gap}^i = g_{gap}(v_m - v_{gap}^i)$$

where $C_m$ is the membrane capacitance, $v_m$ is the membrane voltage, $v_{gap}^i$ is the membrane voltage of neighboring neurons connected by gap junction, $k$ is the number of neighboring neurons, $g_{leak}$ is the conductance of passive leak channel, $e_{leak}$ is the potential of passive leak channel, $g_{gap}$ is the conductance of gap junction, and $I_{photo}$ is the photocurrent transformed from the input.

Each photoreceptor encodes single pixel. The photocurrents are reduced to a double-exponential function:

$$I_{photo} = I_{dark} - g_{max} \frac{\tau_2(e^{-\frac{t}{\tau_1}} - e^{-\frac{t}{\tau_2}})}{(\tau_1 - \tau_2)\frac{\tau_2}{\tau_1}^{\frac{\tau_1}{\tau_1 - \tau_2}}}$$

where $I_{dark} = 40\text{pA}$, $\tau_1 = 64\text{ms}$, $\tau_2 = 68\text{ms}$, and $g_{max} = \text{pixel}/255 * 15.686 \text{ pA}$ is the peak of $I_{photo}$.

### Shallow Retina Block

Shallow Retina Block is a biologically constrained RNN model equivalent to G-filter. SRBlock contains $m * n$ neurons, which has the same size with input image. Each neuron connects to itself and to neighboring neurons only. Here we elaborate on how to derive the RNN implementation. The implementation details of SRBlock can be divided into the next three parts.

### a) pixel-to-sequence converter

Each RNN neuron receives a single pixel $U_{i,j}$ of image U as a photo signal. $U_{i,j}$ will be mapped as a sequence $<x_0, x_1, x_2, \dots x_n>$ of length n before being processed by RNN neuron. The mapping is defined as:

$$f(t) = \frac{\tau_2(e^{-\frac{t}{\tau_1}} - e^{-\frac{t}{\tau_2}})}{(\tau_1-\tau_2)\frac{\tau_2}{\tau_1}^{\frac{\tau_1}{\tau_1-\tau_2}}}$$

$$x_t = x_0 - gU_{i,j}f(t)$$

where $x_0 = 4e-11, g = 1.5686e-4, \tau_1 = 64, \tau_2 = 68$.

### b) weight-predefined RNN

Each RNN neuron only connects to itself and its neighboring neurons in SRBlock. The network weights in RNN have two parts:

1) self-to-self weight $W_l$, which can be denoted as:

$$W_l = G_l I$$

where I is the identity matrix and $G_l = -1.4e^{-9}$.

2) self-to-neighbor weight $W_\alpha$. Here we adopt a hyper-parameter λ to describe the connection strength, then $W_\alpha$ can be denoted as:

$$W_\alpha = \lambda(W_A - W_D)$$

where $W_A$ denotes the adjacency matrix of RNN, and $W_D$ denotes a diagonal matrix, whose main diagonal element in each position is the number of neighbor neuron in this position.

According to Eq.3, at time step k, a single RNN neuron is ready to generate output $y_{k+1}$, which is calculated from three parts: 1) activation from the new input $x_k$, 2) self activation from the last output $y_k$ and 3) gap junction interaction from the last output of neighboring neurons, $p_k$, $q_k$, etc. The number of this kind of connections can be two, three or four according to the different position of neurons.

$z(t)$ in Eq.3 contains two variables $v_m(t)$ and $v_{gap}^i(t)$. To reduce the computational complexity, $v_m(t)$ and $v_{gap}^i(t)$ can be unified as elements in matrix $Y_t$ by considering from the matrix view:

$$Y_t = Y_{t-1} + \alpha((W_\alpha + W_l)Y_{t-1} + X_t + B)$$

$$\alpha = \frac{dt}{C_m}$$

$$B = -G_l E_l$$

where $Y_t$ denotes all neurons' outputs in time step t and $X_t$ denotes all neurons' inputs in time step t, $dt = 5e^{-4}$, $C_m = 3e^{-11}$, $E_l = -0.07I$, $Y_0 = -0.04142868I$ and $I$ is the identity matrix.

### c) sequence-to-pixel converter

At time step t, the output $y_t$ of each neuron is recorded and finally we obtain an output sequence $<y_0, y_1, y_2, ... y_n>$. After taking absolute values and max-pooling of this sequence, we have:

$$V_{i,j} = max\{|y_k|\}$$

where $V_{i,j}$ is the output value of neuron in position $(i,j)$. Here we perform normalization on the output V to meet the requirements of image domain:

$$O = \frac{V - min(V)}{max(V) - min(V)}$$

### d) Recurrent core unfolded by time steps

We can unfold the RNN structure in time to produce the appearance of layers, in which each layer represents the data processing in each time step. In our experiments, we set the length of the input sequence $<x_0, x_1, x_2, ... x_n>$ to 600, which means the unfold RNN can be regarded as a 600 layers' network. Between each two layers, only neurons correspond to the same position or the adjacent location can establish connections. For ablation study, we can truncate the network at layers of different depths. For example, when we set the length of the input sequence to 300, the RNN output is the sequence $<y_0, y_1, y_2, ... y_{300}>$ instead of $<y_0, y_1, y_2, ... y_{600}>$. As we use a max-pooling process behind the weight-predefined RNN (4.2d), the two output value $V^{300} = max_1^{300}\{|y_k|\}$ may be the same as $V^{600} = max_1^{600}\{|y_k|\}$. Our statistics show that most of the peak value (99.99%) locates between layer 120 and layer 240 (named as peak value region). As long as the running time steps exceeds this area, the output value will be the same.

**Training and evaluation details**

In our experiments, we use three typical neural networks, ResNet50[1], VGG19[30], and Inception-V3[50].

ResNet is famous for its shortcut connections, which can hold the input information to next block and learn residual functions. Our implementation for the 50-depth ResNet follows the practice in [53], basic residual blocks are used.

VGG increases the network depth by using an architecture with very small convolution filter. VGG19 contains 16 convolutional layers with 3 × 3 convolution filters and three fully-connected layers. Our implementation follows *https://github.com/chengyangfu/pytorch-vgg-cifar10*.

The strategy of Inception V3 is factorizing and concatenating multiple different sized convolutional filters. We use the network implementation in *https://github.com/weiaicunzai/pytorch-cifar100*.

We used PyTorch 0.4.1 and tensorflow 1.15.0. All experiments not involved in SRBlock are trained and evaluated in PyTorch. And the SRBlock is coded with tensorflow for preprocessing.

We trained our models using the grayed Cifar-10 dataset. The original CIFAR-10 dataset consists of 60000 32x32 images in 10 classes, with 5000 training images and 1000 test images per class. As for the limitation of PR-filter, we do the grayscale before using this dataset. As the input channel is inconsistent with the original models which are designed for color images. We add a convolution layer (input channels=1, output channels=3, kernel size=1) to the top of all three neural networks only for channel expanding.

For training, we use SGD with a mini-batch size of 400 and trained each model on 1GPU(V100) for 90 iterations. We use a weight decay of 0.0001 and a momentum of 0.9, and a cross-entropy loss between image labels and model predictions. The learning rate starts from 0.1 and is divided by 10 every 30 steps. The training profile is the same in all the three networks.

We use basic data augmentation for training: input images (32*32 pixels) are padded with zeros to the size of 40*40, then random cropped to 32*32 pixels and make random flip horizontally. The data transformation was followed by a normalization layer, with a subtraction value 0.481 and a division value 0.239, which are the mean and standard deviation calculated on the grayed Cifar-10 trainingset.

As we regard the G-filter and the SRBlock as preprocesser, they are not evolved in backpropagation. We preprocessed the grayed Cifar-10 dataset using the G-filter or the SRBlock, saved the output with floating point format, and then using these results for model training. The evaluation follows the same filtering and normalization, without other data augmentation.

When salt-and-pepper noise is added for training, each pixel for images in training set became 0(salt) or 255(pepper) with a fixed chance. Then we preprocessed the perturbed images with the G-filter or the SRBlock, and fed to the models. When making evaluation on the models, salt-and-pepper noise is no longer added to the testing set, as it is only used as a way to enhance the robustness of the models after training.

**Adversarial attack and defense**

**a) Adversarial Attacks**

For performing adversarial attacks, we used fast gradient sign method (FGSM), one of the first successful adversarial attack methods. Let $L$ be the differentiable loss function that was used to train the classifier, $I_c$ and $l$ be the original image and its label, $\theta$ be the parameters of the model, $sign()$ be a conversion function only keep the numeric sign of input value, $\varepsilon$ be a configurable step size, then we can calculate the adversarial perturbation $\rho = \varepsilon\, sign(\nabla L(\theta, I_c, l))$. Adding this perturbation to original image would increase the value of the loss function and mislead the deep neural network.

We also used untargeted projected gradient descent (PGD) with $L_\infty$ norm constraints. PGD is a stronger variant of FGSM, which iteratively computes adversarial perturbations $I_\rho$ and projects the values of the pixels within its restriction $\delta$:

$$I_\rho^{i+1} = clip\left\{clip\left\{I_\rho^i + \varepsilon\, sign\left(\nabla L(\theta, I_\rho^i, l)\right), I_\rho^0 - \delta, I_\rho^0 + \delta\right\}, 0, 1\right\}$$

In the above formula, the internal "clip" ensures the final adversarial image is within an $L_\infty$ norm bound around original input $I_\rho^0$. The external "clip" ensures the perturbed sample is within the image domain [0,1].

In FGSM attack, we use a step size of 0.015, which in practice can cause a 3% $L_2$-difference between original images and adversarial ones. In PGD attack, we set the $\varepsilon$ as 0.0012 and $\delta$ as 0.03. The iteration repeats 30 times (also 10, 20, 40, 50, 60, 70, 80, 90, 100 times when make statistics on decision boundary changes in Fig2.E).

The evaluation of adversarial samples follows the same process as [33]. We feed the clean images to models to produce the adversarial samples, use G-filter or SRBlock to make image transformation, and record the accuracy of transformed adversarial samples.

**b) Other defense methods**

We compared the performance of our methods in the described benchmark with three other methods: JPEG compression[32], bit-depth reduction[33], and adversarial training[35].

Bit-depth reduction and JPEG compression are both transformation-based methods. They defense against adversarial attacks by discarding part of image information on the adversarial samples and making compression.

For the former, the bit-depth of image pixels are reduced from 8 bits to 3 bits as follows:

$$Transformed_i = round(\frac{2^N i}{2^N})$$

Where $i$ is the original image in tensor domain($[0,1]$), $N$ is the reduction depth (3 in our paper), function $round()$ rounds a number to a integer.

For the latter, we used the interface from the PIL library to transform images to JPEG format at quality level 75 (out of 100). Both follows the practice in 33. The training, evaluating, and attack process of these two methods is the same as G-filter and SRBlock.

Adversarial training is a classical training-based defense method. The goal is to optimize the following saddle point problem:

$$\min_\theta \rho(\theta), \quad where\ \rho(\theta) = E_{(x\sim y)\sim D}[\max_{\delta \in S} L(\theta, x + \delta, y)]$$

Where $\theta$ is the set of model parameters, $(x\sim y)$ are pairs of examples and corresponding labels in data distribution D. When given the loss function $L(\theta, x, y)$, the above formula tries to minimize the risk of misclassification with a set of small perturbations $S$. In our experiments, D is the grayed Cifar-10 dataset and PGD attack is used to find the perturbation maximizing the loss. In practice, we use $||\delta||_\infty = \frac{8}{255}$, $\varepsilon = \frac{3}{255}$, and 30 times iteration to generate the adversarial samples for training, which follows the work in 35.

### Decision Boundary Analysis
### a) 2D Decision Boundary

We visualize the manifold and its decision boundary of deep neural network by gradients24. Let $L(x, y)$ be the cross-entropy loss when feeding input $x$ and target $y$, we can calculate the gradient $g$ of the loss surface as:

$$g = \nabla L(x, y)$$

First, we decompose $g$ into two perpendicular directions, the smoothed-gradient $G$ and the orthogonal direction $h$. We random sample 10 perturbations $\xi_{1\sim 10}$ from the gaussian noise distribution with $\sigma = 0.02$ and add them to the original input $x$ respectively. The smoothed-gradient $G$ is calculated as the mean gradient of these 10 samples:

$$G = \frac{1}{10}\sum_{i=1}^{10} \nabla L(x + \xi_i, y) \quad \xi_i \sim N(0, \sigma^2)$$

Another direction $h$ is obtained by decomposing $g$, while ensuring that $h$ and $G$ are orthogonal:

$$\hat{G} = \frac{G}{||G||_2}$$

$$h = g - <g, \hat{G}> \hat{G}$$

$$\hat{h} = \frac{h}{||h||_2}$$

Here $\hat{G}$ and $\hat{h}$ denote the unit vectors in these two directions.

Each point $(u, v)$ in the 2-D decision boundary plane represents the original image $x$ perturbed by $u$ and $v$ along these two directions, clipping as $(x + u\hat{G} + v\hat{h}, 0, 1)$, and the color corresponds to the label predicted by the neural network. In practice, the max and min value of $u$ and $v$ is $\pm 0.4$, while the grid step is 0.02. The whole 2D Decision Boundary consists of 1681(41*41) samples' prediction results.

When we analyze the decision boundary of transformation-based defense methods (includes G-filter, Bit-depth and JPEG compression) and adversarial training, we first preprocess the testing set with these methods. The transformed images are used as the model input.

### b) The correct prediction region in 2D Decision Boundary

We consider the correct prediction region as the largest connected block of correct labels in the 2D plane. Breath First Search(BFS) is used to find this region. Other prediction results inside the region are regarded as part of the region. The contour C of this region, which is the set of pixels in the region adjacent to points outside the region, is used for the following two measurement. For convenience, we use pair of integers $(u_i, v_i)$ to represent the point i in the 2-D decision. For example, $(0, 0)$ represents the original image $x$ and $(-20, -20)$, $(20, 20)$ represent two farthest points on the plane.

### c) The distance from the gravity center of the correct prediction region's contour to the zero point

Let $n$ be the number of points in the set C, then

$$f = (\frac{1}{n}\sum x_i, \frac{1}{n}\sum y_i) \quad (x_i, y_i) \in C$$

$$d = ||f||_2$$

where $f$ denotes the center of gravity of the contour C, $d$ denotes the distance we calculated.

If there is no correct prediction sample in all the 1681 samples, which means the correct prediction region does not exist, then the gravity distance is calculated as if the gravity locates on the farthest point of the plane, producing a distance of $||(20,20)||_2$

### d) The dispersion of the correct prediction region's contour

The dispersion of the correct prediction region's contour is calculated as:

$$\sqrt{\frac{1}{n}\sum(\sqrt{x_i^2 + y_i^2} - \frac{1}{n}\sum\sqrt{x_i^2 + y_i^2})^2} \quad (x_i, y_i) \in C$$

which is the standard deviation of the distance between each correct label from zero point in the set C. If the correct prediction region does not exist, we record the value 0 for statistics.

### Local Linearity Analysis

Let $f(x)$ be the mapping of the DNN from the input to the classification score(output before the softmax) of the true label. When we use FGSM to attack the DNN and add the adversarial perturbation ε to the original image $x$, we can get a new score $f(x+\varepsilon)$. Due to the destructive nature of the attack, $f(x+\varepsilon)$ is usually smaller than $f(x)$. By continuously increasing the attack strength of FGSM, we can draw the curve of $m(\varepsilon) = f(x+0) - f(x+\varepsilon)$, with the horizontal coordinate be the step size of FGSM (which is the same as $||\varepsilon||_\infty$).

Let $\varepsilon^*$ be the minimal norm perturbation of misclassification when increasing the attack strength of FGSM. We can trace a line $n(\varepsilon) = k\varepsilon$ between the zero point and the point $(2\varepsilon^*, f(x) - f(x+2\varepsilon^*))$. If DNNs have strong local linearity, the error between $m(\varepsilon)$ and $n(\varepsilon)$ will approximate zero between $[0, 2\varepsilon^*]$. Therefore, we denote the mean square error N between $m(\varepsilon)$ and $n(\varepsilon)$ to be the measurement of nonlinearity.

$$N = \sqrt{\sum_{\varepsilon=0}^{2\varepsilon^*}(m(\varepsilon) - n(\varepsilon))^2}$$

In practice, we increase the step size of FGSM from 0 to 1 at the intervals of 0.001 and make records during the 1000 attacks. Clip will be used when the step size is too large to keep $x+\varepsilon$ within the image pixel domain $[0,1]$. If $||[\varepsilon^*]||_\infty$ is larger than 0.5, which means $||[2\varepsilon^*]||_\infty$ will be larger than 1, we take the farthest point of $m(\varepsilon)$, whose abscissa is 1, as an alternative of $m(2\varepsilon^*)$ and $n(2\varepsilon^*)$.

When we measure the nonlinearity of transformation-based defense methods (includes G-filter, Bit-depth and JPEG compression) and adversarial training, we follow the same method to generate the FGSM samples. The transformed adversarial samples are used to calculate the score $f(x+\varepsilon)$ and to determine whether a misclassification has occurred.

**Worst-case Robustness Analysis**

We use the deepfool algorithm to iteratively calculate the worst case robustness following the protocol in 40,42,51. For iteration i, we approximate the true decision boundary by the complement of a convex polyhedron and calculate the perturbation vector that reaches each binary classification boundary of the polyhedron. Sample $x_{i-1}$ is then pushed to the nearest boundary and produce a new perturbed sample $x_i$. We continue to do this until the perturbed image crosses over the true decision boundary and misclassifies the model.

Let $t$ be the label of image $x_0$, $f_k(x)$ be the model output (before the softmax) at channel $k$ ($k$ is an integer between 1 and 10 corresponding to 10 classes). Traverse all k values that are not equal to t and calculate:

$$w'_k \leftarrow \nabla f_k(x_i) - \nabla f_t(x_i)$$
$$f'_k \leftarrow f_k(x_i) - f_t(x_i)$$
$$l_k \leftarrow \frac{|f'_k|}{||w'_k||_2}$$

The shortest perturbation vector at iteration i is calculated as:

$$k^* \leftarrow argmin_{k \neq t} l_k$$

$$r_i \leftarrow \frac{|f'_{k*}|}{||w'_{k*}||_2^2} w'_{k*}$$

The perturbation accumulated from the first iteration is:

$$\rho_i = \sum_1^i r_i$$

New perturbed image $x_i = x_0 + (1 + overshoot)\rho_i$. Here we set the overshoot to 0.02 and the maximum iteration to 100. If the perturbed image can successfully mislead the model within 100 iterations, for example, at iteration n, then the worst-case robustness R is measured by $||\rho_n||_2$. If this is not the case, we use $||\rho_{100}||_2$ as an approximation.

During the iteration, we record the accumulated perturbation $<\rho_1, \rho_2, ...\rho_n>$. Then for each iteration i, we can calculate

$$R_i = ||\rho_{min\,(n,100)} - \rho_i||_2$$

Which denotes the worst-case robustness of the original image after i iterations.

When we measure the worst-case robustness of transformation-based defense methods (includes G-filter, Bit-depth and JPEG compression) and adversarial training, we feed the clean images to models and apply deepfool algorithm as the same. The difference is that the transformed samples $x_i^{trans}$ instead of $x_i$ are used to determine whether a misclassification has occurred.

**Local Linear Embeddding Analysis**

Local linear embedding (LLE) algorithm is a high-dimension to lower-dimension projection method which preserves distances within local neighborhoods[52]. We use LLE to project the direction of the smooth gradient in each PGD iteration into the 3D space.

We random select 1000 samples from the testing set. From PGD iteration 1 to iteration 30, we feed the generated adversarial images to the model and record the normalized smooth gradient $G$ (described in Decision Boundary Analysis). The data space shows in the form of 30000(30*1000) points in 32*32 dimensions. We apply LLE method on this data space and reduce the dimensions to 3. For the new vectors of size(30,1000,3), we calculate the mean value of each iteration and get 30 3-D vectors representing the reduced smoothing gradient after each iteration. We connect them end to end to show the general trend.

The LLE code is from https://cs.nyu.edu/~roweis/lle/code.html and 28 nearest neighbors are used.

When applying G-filter, we feed the transformed adversarial images to record the normalized smooth gradient $G$, following the same procedure in Decision Boundary Analysis.